\documentclass[twocolumn,aps,showpacs,preprintnumbers,lettersize]{revtex4}
\usepackage{graphicx}
\usepackage{dcolumn}
\usepackage{bm}
\usepackage{color}
\usepackage{multirow}

\def\ga{{\ \lower-1.2pt\vbox{\hbox{\rlap{$
          >$}\lower5pt\vbox{\hbox{$\sim$}}}}\ }}
\def\la{{\ \lower-1.2pt\vbox{\hbox{\rlap{$
          <$}\lower5pt\vbox{\hbox{$\sim$}}}}\ }}
\def\beq{\begin{equation}}
\def\eeq{\end{equation}}
\def\bea{\begin{eqnarray}}
\def\eea{\end{eqnarray}}
\def\he3he4{$^3$He-$^4$He}
\def\eps{\varepsilon}

\def\cost2e{\cos\frac{\theta}{2}}
\def\cost2t{\cos\theta/2}
\def\he3{$^3$He}
\def\he4{$^4$He}

\def\g1t{$\tilde{\Gamma}_1$}

\definecolor{og}{named}{OliveGreen}

\begin{document}

\title{Effect of superconducting fluctuations on ultrasound in unconventional
superconductor}\author{M.S.~Mar'enko$^{1}$}\author{C.~
Bourbonnais$^{1,2}$}\author{A.-M.S.~Tremblay$^{1,2}$}
\affiliation{$^{1}$D\'{e}partement de physique and Regroupement
qu\'{e}b\'{e}cois sur les mat\'{e}riaux de pointe, Universit\'{e}
de Sherbrooke, Sherbrooke, Qu\'{e}bec, J1K 2R1,
Canada\\$^{2}$Institut canadien de recherches avanc\'{e}es,
Universit\'{e} de Sherbrooke, Sherbrooke, Qu\'{e}bec, J1K 2R1,
Canada }\date{\today }

\begin{abstract}
We study the renormalization of sound attenuation and sound
velocity by fluctuation Cooper pairs in layered superconductors.
We consider the influence of $s$- and $d$-wave symmetry of the
fluctuating order parameter, on both longitudinal and transverse
phonon modes. We show that both unconventional order parameter
symmetry and transverse sound polarization suppress the AL and MT
terms, while the DOS contribution is the least affected.  The
combination of these effects can change the sign of the overall
fluctuation corrections above $T_c$. We also compare the results
obtained using the Ginzburg-Landau formalism with a microscopic
derivation of the fluctuation corrections to  the sound velocity
   in both $s$- and $d$-wave superconductors.
These calculations are motivated by ongoing ultrasound
measurements in organic superconductors.
\end{abstract}\maketitle

\section{Introduction}

Ultrasound is a very powerful tool to study the electronic
properties of metals. In systems such as Sr$_2$RuO$_4$,\
\cite{Louis_Sr2RuO4} heavy fermions,\
\cite{usound_ex_heavy_fermions} and organic superconductors\
\cite{usherb_ex} the ultrasound technique was used to probe the
symmetry of the order parameter. In layered organic
superconductors, ultrasound attenuation was one of the key methods
to determine the phase diagram.\ \cite{usherb_ex} Compounds from
the $\kappa$-(ET)$_{2}$-X family can exist in different phases,
such as normal metal, superconductor or Mott insulator. Most phase
transition lines can be associated with anomalies of the sound
velocity and sound attenuation as a function of temperature or
pressure.\ \cite{usherb_ex} In particular, preliminary experiments
indicate that the N-S phase transition is accompanied by a
pronounced fluctuation region. This can be explained as an effect
of transient Cooper pairs above $T_c$, or, equivalently,
fluctuations of the superconducting order parameter. It is known
that in layered organic materials the fluctuation region is
especially broad due to strong electronic anisotropy, low charge
carrier concentration and relatively high $T_c$. The manifestation
of superconducting fluctuations in various physical phenomena in
organic superconductors has been experimentally confirmed.\
\cite{fluct_ex, KFrikash}

The sound attenuation and the sound velocity in metals are
determined at low enough temperatures  by the interaction of
phonons with the electronic system. In the Fr\"{o}hlich model,
phonons couple to the electronic density, hence the
 electron-phonon interaction is momentum-independent. This approach is
not valid,\ \cite{Schmid} however, in many systems of interest. In
the presence of impurities, one has to consider the electron
system in a reference frame moving together with the ion lattice.\
\cite{Schmid, Pippard, Tsuneto} Indeed, in an impure metal, the
elastic scattering of electrons as well as perfect screening at
small phonon momentum and relaxation to equilibrium occur in this
oscillating frame. In this formalism, the electron-phonon coupling
appears through the stress tensor, rather than through the density
operator.

In order to include polarization effects on sound propagating in
highly anisotropic media such as organic superconductors, one has
to consider a tight-binding model instead of a continuous one. In
the tight-binding model, the electron-phonon interaction comes
from the modulation of the hopping parameter induced by the
lattice deformation when the sound wave propagates through the
crystal. The electron-phonon vertex then depends on the
orientation of the phonon momentum and the phonon polarization. In
such a model, where the stretching of specific crystal bonds
induced by the lattice deformation contributes to the
electron-phonon vertex,\ \cite{SamokhinWalker} one can explain for
example that anisotropy in the attenuation of different phonon
modes can be as large as few orders of magnitude in
Sr$_2$RuO$_4$.\ \cite{Louis_Sr2RuO4}

In the superconducting state, the symmetry of the order parameter
can also affect the temperature dependence of the sound
attenuation. Namely, it does depend on the presence of nodes and
on their orientation with respect to the phonon momentum and
polarization. Thus, sound attenuation experiments can probe the
type of superconducting pairing.\ \cite{ColemanMoreno,
Contreras04}

One would naturally expect that the superconducting fluctuations
\cite{AL, Maki,Thompson}
 will interact with ultrasound in a
similar way, thus revealing the order parameter symmetry. The
magnitude of the ultrasound renormalization by fluctuations is not
necessarily small. Previously, the fluctuation corrections to
longitudinal phonon mode propagating perpendicular to the
conduction layers in quasi-2D material were found assuming
$s$-wave symmetry of the energy gap.\ \cite{MBT04} The model that
uses the quasi-2D open Fermi-surface and the electron-phonon
vertex originating from the modulation of the electron interlayer
hopping integral in the presence of the ultrasound wave predicts
rounding of the sound attenuation and sound velocity temperature
dependencies near $T \rightarrow T_c$.

Since there is a substantial experimental interest in this
subject, we present estimations for the fluctuation corrections to
the sound attenuation and sound velocity for longitudinal and
transverse ultrasound propagating perpendicular to the conduction
planes, for $s$- and $d$-wave symmetries of the order parameter.

The paper is organized as follows: in Section \ref{sec_thermo} we
present a simple estimation for the fluctuation corrections to the
sound velocity that can be obtained from the Ginzburg-Landau
formalism in order to provide a simple phenomenological basis for
the microscopic results.  In Section \ref{secModel} we describe
the microscopic model for the quasi-2D superconductor and for the
electron-phonon interaction at various sound polarizations. In
Section \ref{sec_no_fluctuations} we review the sound attenuation
and sound velocity in the mean field approximation (without
superconducting fluctuations). In Sections \ref{sec_fluct_s-wave}
-- \ref{sec_dwave} we give the superconducting fluctuation
corrections for various pairing symmetries and phonon
polarizations. Finally, we present a discussion of our results and
of their relevance to experiment.

\section{Ginzburg-Landau approach to the fluctuation sound velocity\label{sec_thermo}}

In this section, we obtain the fluctuation corrections to the
sound velocity from the Ginzburg-Landau free-energy functional.
Before that, we shall give, for completeness, a summary of
elasticity theory  that can be found in more details elsewhere.\
\cite{LLvol7, AshcroftMermin} Even though it is less detailed than
the microscopic approach that is given later, the Ginzburg-Landau
point of view allows one to develop physical intuition. It can
also provide an independent check of the microscopic results that
are given in the rest of the paper.

\subsection{Sound velocity jump at $T=T_c$ \label{subsec_Testardi}}

The difference between superconducting and normal state free
energies at temperatures close to $T_c$ without fluctuations is
\beq\label{free_energy_difference} \Delta F_{NS} =  F_{N} -  F_S =
\frac{\alpha^2 [T-T_c(\epsilon_{ij})]^2 }{8 \pi}, \eeq where
$\alpha$ is the Ginzburg-Landau coefficient, and
\beq\label{strain_definition} \epsilon_{ij}=\frac{1}{2}
\left(\frac{\partial u_i}{\partial x_j} + \frac{\partial
u_j}{\partial x_i}\right)\eeq is the (3$\times$3) strain tensor
that is defined through the displacement ${\bf{u}}$ at a point
${\bf{x}}$. In a continuous elastic medium, $\epsilon_{ij}$ obeys
Hooke's law: \beq\label{Hookes_law} \sigma_{ij} = c_{ijkl}
\epsilon_{kl},\eeq where $\sigma_{ij}$ is the stress tensor and
$c_{ijkl}$ the elastic modulus tensor. Symmetry considerations
allow us to reduce the number of independent components in
 Eq. (\ref{Hookes_law}), thus lowering the rank of $\epsilon_{ij}$ and
$\sigma_{ij}$ and converting them into (1$\times$6) vectors
$\epsilon_{m}$ and $\sigma_{m}$ with the following rule that
relates the indices $\{ij\}$ and $\{m\}$:
\beq\label{conversion_rule} \!\! xx\rightarrow 1,\ yy\rightarrow
2,\ zz\rightarrow 3,\ yz\rightarrow 4,\ zx\rightarrow 5 ,\
xy\rightarrow 6.\eeq

Similarly, the elastic modulus tensor $c_{ijkl}$ is converted into
a symmetric $(6\times 6)$ tensor $c_{mn}$ according to
$ji\rightarrow m$, $kl\rightarrow n$, and the rule
(\ref{conversion_rule}). The tetragonal system symmetry reduces
the number of independent components of $c_{mn}$ to only 6 and
imposes additional relations between them: $c_{11}=c_{22}$,
$c_{44}=c_{55}$, $c_{12}=c_{21}$, $c_{13}=c_{23}=c_{31}=c_{32}$,
and all other off-diagonal components are 0.

The equation of motion for  sound waves results in the relation
for the sound velocity $v_s$: \beq\label{sound_velocity_density}
{\bar{c}}=\rho {v^2_s}, \eeq where $\rho$ is the density, and
${\bar{c}}=c_{ijkl} \hat{q}_i \hat{q}_j  \hat{e}_k \hat{e}_l$ is
the contraction of the elastic modulus tensor with the phonon
momentum and polarization vectors. In particular, for longitudinal
phonons propagating along $z$-axis, the ${\bar{c}}$ being
expressed in terms of the $(6\times 6)$ tensor is
${\bar{c}}=c_{33}$. For transverse phonons with the momentum
${\widehat{\bf{q}}}||\widehat{{\bf{z}}}$ and polarization
$\widehat{{\bf{e}}}$ lying in the $xy$ plane, ${\bar{c}}=c_{13}$
for any angle $\varphi$ between the vector $\widehat{{\bf{e}}}$
and $x$-axis.

Then, with the notations \beq\label{GammaDelta}
\Gamma_m=\frac{\partial T_c}{\partial \epsilon_{m}},
~~~\Delta_{mn}=\frac{\partial^2 T_c}{\partial \epsilon_{m}
\partial \epsilon_{n}},\eeq the discontinuity in the elastic
modulus tensor at $T=T_c$ that is related to the second derivative
of $\Delta F_{NS}$  with respect to strain is \cite{Testardi}:
  \beq \Delta c_{mn}=\frac{1}{V_0} \left.
\frac{\partial^2 \Delta F_{NS}}{\partial \epsilon_{m}
\partial \epsilon_{n}} \right|_{T=T_c}=\frac{\alpha^2
\Gamma_m\Gamma_n}{4\pi V_0}, \eeq where $V_0$ is the unstrained
volume. We shall consider the thermodynamics of the longitudinal
phonons only, because for transverse modes $\Gamma_m$ vanishes by
symmetry hence there is no discontinuity in $c_{mn}$. Using the
relation for the sound velocity Eq.
(\ref{sound_velocity_density}), one finally obtains the sound
velocity jump at $T=T_c$: \beq \Delta {v_s}= \frac{\Delta
{\bar{c}}}{2 v_s \rho} = \frac{\alpha^2 {\overline{\Gamma_m
\Gamma_n}}}{8 \pi v_s \rho V_0}, \eeq where the summation in
${\overline{\Gamma_m \Gamma_n}}$   is done over all indices $m$
and $n$ that correspond only to non-zero components of ${\bar{c}}$
for the particular ${\widehat{\bf{q}}}$ and
${\widehat{\mathbf{e}}}$. Finally, the sound velocity jump can
also be related to the specific heat jump at $T=T_c$ \beq \Delta
C_{NS}=-T_c \frac{\alpha^2}{4 \pi V_0} \eeq as follows:\
\cite{Testardi} \beq\label{vs_c_relation} \frac{\Delta
{v_s}}{v_s}=-\frac{\Delta C_{NS}}{T_c} \frac{1}{2 \rho}
\overline{\left( \frac{1}{v_s} \frac{\partial T_c}{\partial
\epsilon_{m}} \right) \left( \frac{1}{v_s} \frac{\partial
T_c}{\partial \epsilon_{n}} \right)}. \eeq This relation is quite
general. It holds for both isotropic and anisotropic materials in
the mean field approximation.

\subsection{Fluctuation corrections to the sound velocity at $T>T_c$ in the 3D isotropic case \label{subsec_GL_v_3D}}

We can use the Ginzburg-Landau formalism to obtain the corrections
to the sound velocity at $T>T_c$. In the presence of
superconducting fluctuations, the free energy acquires the term :
\beq\label{Omega_fl} F_{\mbox{fl}} =-T \sum_{p<p_0} \ln \frac{\pi
}{\alpha \left(\epsilon + \eta_c{ p^2} \right)}, \eeq where
$\epsilon=(T-T_c)/T$, $\eta_c=1/(4 m \alpha T  )$, and $p_0\sim
1/\sqrt{\eta_c}$ is a cut-off. Then the most singular
contributions are: \beq\label{ddOmega} \frac{\partial^2
F_{\mbox{fl}}}{\partial \epsilon_{m}\partial \epsilon_{n}} =-
\sum_{p<p_0} \left[ \frac{ \Gamma_m\Gamma_n}{ T \left(\epsilon +
\eta_c { p^2}\right)^2} + \frac{ \Delta_{mn}}{\epsilon + \eta_c {
p^2}}\right].\eeq Here the structure of the second term resembles
the 3D fluctuation propagator $L(\Omega=0,p)$ in the Gaussian
fluctuation theory \cite{LV01}, while the first term of Eq.
(\ref{ddOmega}) looks like the fluctuation propagator squared.

In Eq. (\ref{ddOmega}), the main contribution is from the first
term. If one changes summation to integration and sets the upper
limit to infinity, in the limit $\epsilon\rightarrow 0$ it reads:
\beq\label{Lsquared} - \int_0^{p_0} \frac{ \Gamma_m\Gamma_n}{ T
\left(\epsilon + \eta_c { p^2}\right)^2} \frac{2 \pi p^2 }{(2
\pi)^3} dp = -\frac{\Gamma_m\Gamma_n }{16 \pi T_c \eta_c^{3/2}
\sqrt{\epsilon}}.\eeq This expression is singular at $T=T_c$ and
clearly resembles the standard fluctuation conductivity
contribution of the Aslamazov-Larkin type in the 3D case
\cite{AL}. The second term of Eq. (\ref{ddOmega}) being integrated
over momenta $p$ is not singular as $\epsilon \rightarrow 0$.

Thus, considering for simplicity only diagonal terms of the strain
tensor, the corrections to the sound velocity are given by:
\beq\label{vs_fluct_3D} {v_s}_{\mbox{fl}} = \frac{1}{2 v_s \rho }
\frac{\partial^2 F_{\mbox{fl}}}{\partial \eps_{m}\partial
\eps_{m}}=-\frac{\Gamma_m\Gamma_n }{32 \pi v_s \rho T_c
\eta_c^{3/2} \sqrt{\epsilon}}. \eeq

Note that the standard derivation of the fluctuation corrections
to the electronic specific heat, that requires taking second
derivative of the thermodynamic potential with respect to
temperature, results in the expression\ \cite{Levanyuk}\beq
C_{\mbox{fl}}=\frac{ 1 }{16 \pi \eta_c^{3/2} \sqrt{\epsilon}},
\eeq that is in agreement with  Eqs. (\ref{vs_c_relation}) and
(\ref{vs_fluct_3D}). The fluctuation corrections to sound velocity
and specific heat here are connected in the same way as the
mean-field jumps. This is not the case in the Lawrence-Doniach
model, as we will discuss in the next section.

\subsection{Fluctuation corrections in Lawrence-Doniach model \label{subsec_Lawrence-Doniach}}

In the case of layered materials, one should use the
Lawrence-Doniach quasi-2D free energy rather than the isotropic 3D
one. At a low electron density, the quasiparticle spectrum can be
taken to be
\begin{equation}\label{qparticle_spectrum_lowdensity}\xi
(p_{||},p_{z})=\frac{p_{||}^{2}-p_{F}^{2}}{2m}-2t_{\perp }\cos {(p_{z}c)}
\label{spectrum}\end{equation} where $p_{||}$ denotes the
intralayer component of the momentum, $p_z$ the interlayer
component, $c$ the interlayer distance, and $t_{\perp }$ the
interlayer hopping integral. The Fermi surface then has the form
of a corrugated cylinder with the one-particle density of states
given by $\nu_{0}=m/(2\pi c)$. Then, the fluctuation term
analogous to Eq. (\ref{Omega_fl})   reads:
\beq\label{Omega_fl_q2D} F_{\mbox{fl}} =-T \sum_{p<p_0} \ln
\frac{\pi }{\alpha \left\{ \epsilon + \eta_{||} { p_{||}^2}
+\frac{r}{2} [1-\cos(p_z c)]\right\}}. \eeq Now $\eta_{||}$ is
related to the square of the in-plane correlation length, and the
parameter $r $ characterizes the material anisotropy. The
modulation of hopping integral $\delta t_\perp$ by the strain
$\epsilon_m$ due to the lattice deformation by the sound wave
contributes to the electron-phonon interaction, that should be
taken into account in the second derivative of $F_{\mbox{fl}}$ Eq.
(\ref{Omega_fl_q2D}) with respect to strain. Without going into
all the details, we assume for the moment simple proportionality
relations: $r\propto t_\perp^2$ and $\delta t_\perp \propto
(\partial t_\perp / \partial u_i) (\partial u_i / \partial x_j)
\delta x_j$, where $(\partial t_\perp / \partial u_i)$ contributes
to the appropriate electron-phonon coupling constant $g$, and the
combination $(\partial u_i / \partial x_j)$ can be rewritten in
terms of the strain tensor Eq. (\ref{strain_definition}).
Therefore, one has: \bea\label{LD_qint} &&\frac{\partial^2
F_{\mbox{fl}}}{\partial \epsilon_{m}\partial \epsilon_{n}}
=-\frac{1}{T}
\int_0^{p_0} \int_{-\pi/c}^{\pi/c}\frac{dp_z p_{||}dp_{||} }{(2 \pi)^2}\\
&& \times \frac{\{ \Gamma_m  - g t_\perp T [1-\cos(p_z c)]\} \{
\Gamma_n  - g t_\perp T [1-\cos(p_z c)]\} }{\left\{\epsilon +
\eta_{||} { p_{||}^2}+\frac{r}{2} [1-\cos(p_z
c)]\right\}^2} \nonumber\\
&& -  \int_0^{p_0} \int_{-\pi/c}^{\pi/c} \frac{dp_z p_{||}dp_{||}
}{(2 \pi)^2}\frac{ \Delta_{mn}- g^2  T [1-\cos(p_z c)]}{\epsilon +
\eta_{||} { p_{||}^2}+\frac{r}{2} [1-\cos(p_z c)]}.\nonumber \eea

We separate the expression in the last line of Eq. (\ref{LD_qint})
into two parts. The first one, without $\cos(p_z c)$ in the
numerator, results in the following contribution to the sound
velocity: \beq\label{LD_DOS} -\frac{\nu_0 (\overline{\Delta_{mn}}
-g^2 T)}{2 m v_s \rho \eta_{||} } \ln
\left(\frac{2}{\sqrt{\epsilon}+\sqrt{\epsilon+r}} \right).\eeq The
other part, with $\cos(p_z c)$ in the numerator, gives a
contribution of the type: \beq\label{LD_rMT} -\frac{g^2 T \nu_0}{4
m v_s \rho \eta_{||} } \frac{\left(
\sqrt{\epsilon}-\sqrt{\epsilon+r}\right)^2}{r}. \eeq

In the integral in the second line of Eq. (\ref{LD_qint}), one can
keep the terms in the numerator that do not contain elastic
derivatives $\Gamma_m$. This leads to a contribution to the sound
velocity of the following type: \beq\label{LD_AL_tight} -\frac{g^2
T \nu_0}{2 m v_s \rho \eta_{||} }\left[1+ \frac{2 \epsilon}{r}
\left(-1+\sqrt{\frac{ \epsilon}{\epsilon+r}}\right)\right] .\eeq

Keeping the $\Gamma_m \Gamma_n$ term in the numerator of the
integral in the second line of Eq. (\ref{LD_qint}), and neglecting
the hopping terms, gives the contribution:
 \beq\label{LD_AL_cont}
-\frac{\nu_0 \overline{\Gamma_m\Gamma_n}}{4 m v_s \rho T_c
\eta_{||} \sqrt{\epsilon(\epsilon+r)}}. \eeq

Finally, the cross-term that is proportional to both $\Gamma_m$
and $ g t_\perp$, results in the contribution:
\beq\label{cross-term}-\frac{g \nu_0
\overline{(\Gamma_m+\Gamma_n)} }{2 m v_s \rho   \eta_{||}}
\frac{\left(
\sqrt{\epsilon}-\sqrt{\epsilon+r}\right)}{\sqrt{r(\epsilon+r)}}
\eeq that is less singular than Eq. (\ref{LD_AL_cont}) at small
$\epsilon$ and $r$.

Note that the relation (\ref{vs_c_relation}) between the
mean-field jumps of the sound velocity and specific heat at
$T=T_c$ is satisfied in the Lawrence-Doniach model. This comes
about from the fact that in this model the difference between
superconducting and normal state free energies is still given by
Eq. (\ref{free_energy_difference}) with the parameters modified
for the quasi-2D case. On the other hand, the relation Eq.
(\ref{vs_c_relation}) is not satisfied for the fluctuation
corrections in this model. This is because we can neglect any $T$
dependence of the interlayer hopping, hence it does not contribute
to the specific heat. We will obtain later in the microscopic
calculations in Section \ref{sec_fluct_s-wave} the expressions
that have the temperature dependence of  Eqs.
(\ref{LD_DOS})-(\ref{LD_AL_tight}).

\section{Microscopic model\label{secModel}}

\subsection{Quasiparticle energy spectrum\label{subsecSpectrum}}

We use the model of quasi-2D square lattice with the in-plane
period $a$, and the interlayer distance $c>a$. The quasiparticle
energy spectrum in the normal state in the lattice model is given
by: \bea\label{qparticle_spectrum_tight}
&&\xi({\mathbf{p}})=-2 t_{||} [\cos (p_x a) + \cos (p_y a)]-\mu\nonumber\\
 &&~~~~~~~~~-2 t_\perp \cos (p_z c).\eea
 Here $t_{||}$ and $t_\perp$ are the intralayer and interlayer hopping integrals.
 In highly anisotropic materials we have $t_{||}\gg t_\perp$, and the electrons are moving preferentially in
the conduction layers. (In organic materials such as $\kappa
$-(ET)$_{2}$Cu(NCS)$_{2}$, for example, we can estimate the ratio
$t_{||}/t_{\perp }\approx 4000$ \cite{Singleton}). Such a high
anisotropy can raise the issue of incoherent electron motion in
the perpendicular, $z$-direction.\
\cite{Timusk_organics,MosesMcKenzie}
 In the case of weakly incoherent
interlayer motion,\ \cite{MosesMcKenzie} the intralayer electron
momentum is conserved in the tunneling process and the electron
wave function  in adjacent layers has some overlap, but there are
many in-plane collisions between tunneling events. In this case as
well as for coherent interlayer motion, the use of the
quasiparticle spectrum Eq. (\ref{qparticle_spectrum_tight}) is
legitimate. In the case of strongly incoherent interlayer
transport, however, the intralayer electron momentum is not
conserved in the tunneling processes between adjacent layers
because the tunneling can be accompanied by strong elastic or
inelastic processes. This case needs a separate treatment.

The tight-binding spectrum Eq. (\ref{qparticle_spectrum_tight})
for $a p_F \ll 1$ reduces to the form Eq.
(\ref{qparticle_spectrum_lowdensity}). In most cases considered
here the use of the low-density spectrum Eq.
(\ref{qparticle_spectrum_lowdensity}) in the microscopic
calculations is well justified.

\subsection{Electron-phonon vertex\label{subsecGamma}}

The electron-phonon part of the Hamiltonian, according to Walker,
Smith, and Samokhin\ \cite{SamokhinWalker}, is given by
 \bea
&&H_{e-ph}=-\sqrt{2}i\sum_{\mathbf{k,p,R}}G_{\mathbf{R}}\left(
\frac{\hbar \omega _{0}\left(
\mathbf{k}\right) }{NMv_{s}^{2}}\right) ^{1/2}\\
&&\times\left( \widehat{\mathbf{k}} \mathbf{\cdot
}\widehat{\mathbf{R}}\right) \left( \widehat{\mathbf{e}}
\mathbf{\cdot }\widehat{\mathbf{R}}\right) \left( \cos
{\mathbf{p}}\mathbf{\cdot }{\mathbf{R}}\right) \text{
}c_{\mathbf{p}+\mathbf{k},\sigma }^{\dagger }c_{\mathbf{p,}\sigma
}\left(a_{- \mathbf{k}}^{\dagger }+a_{\mathbf{k}}\right) ,
\nonumber \label{He-ph}\eea where $\widehat{\mathbf{k}}$ is a unit
vector in the direction of the phonon momentum,
$\widehat{\mathbf{e}}$ a unit vector for the phonon polarization,
${\mathbf{R}}$ are the nearest bonds that are stretched by the
sound wave. We have also defined $\omega _{0}\left(
\mathbf{k}\right) =v_{s}k$ the sound frequency, $v_{s}$ the sound
velocity, $M$ the ion mass, $N$ the number of unit cells,
$G_{\mathbf{R}} $ a constant that depends on the derivative of the
hopping integral along the bond $\bf{R}$ with respect to the
strain, and finally $a_{\mathbf{k}}^{\left(\dagger \right) }$ and
$c_{\mathbf{p},\sigma }^{\left( \dagger \right) }$ are,
respectively, destruction and creation operators for phonons and
for electrons of spin $\sigma $.
\begin{figure}[h]
\includegraphics[width=8.0cm]{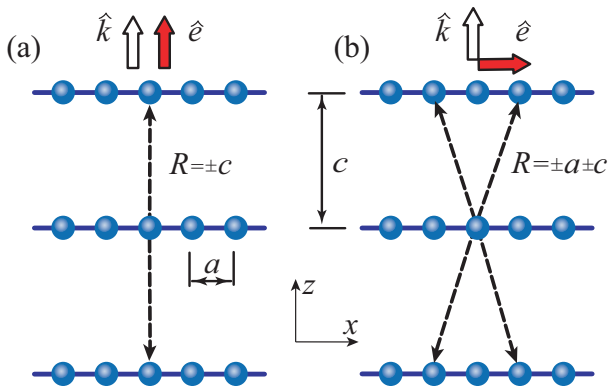}
\vspace{0.8cm} \caption{(Color online) Side view of the basis
vectors that determine the electron-phonon vertex for (a)
longitudinal (L), and (b) transverse (T)  phonons propagating
perpendicular to the layers. Here  $a$ is the in-plane lattice
period, $c$ the interlayer distance, and $\mathbf{R}$ are the
nearest neighbor distances for the bonds that give the leading
contribution to the electron-phonon vertex, for both types of
polarization. }\label{vectors}\end{figure}

For longitudinal phonons propagating in the perpendicular
direction, the main terms in the sum over $\mathbf{R}$ in
 Eq. (\ref{He-ph}) are those that contain the nearest neighbor
interlayer bonds $\mathbf{R}=\pm \mathbf{c}$ (see Fig.
\ref{vectors}(a)). For these bonds, the wave vector and the phonon
polarization will satisfy $(\widehat{\mathbf{k}}\mathbf{\cdot
}\widehat{\mathbf{R}})(\widehat{\mathbf{e}}\mathbf{\cdot
}\widehat{\mathbf{R}})=1$. The electron-phonon vertex  becomes
\beq \Gamma_{ep}^{(L)}({\bf{p}})=g_L\cos(p_{z}c), \eeq where $g_L$
is a constant.

For transverse phonons, stretching of the nearest neighbor bonds
Fig. \ref{vectors}(a) does not contribute to the sound attenuation
as for these bonds
$(\widehat{\mathbf{e}}\mathbf{\cdot}\widehat{\mathbf{R}})=0$. In
other words, the shear wave does not interact with electrons to
this order. One should consider next-to-nearest neighbor bonds,
that in our particular case are the diagonals of the sides of 3D
$a\times a\times c$ unit cell. These bonds belong to the set
$\{(\pm a,0,\pm c), (0,\pm a,\pm c) \}$. Let us assume that the
phonon polarization makes an angle $\varphi$ with the $x$-axis in
the plane, such that $\widehat{\mathbf{e}}=\{\cos\varphi,\sin
\varphi,0 \}$. The summation over the next-to-nearest neighbor
bonds results in an electron-phonon vertex of the form
\beq\label{dt_elph_vertex}\Gamma_{ep}^{(T)}{(\varphi,{\bf{p}})}=g_T
[\sin(p_x a) \cos\varphi+\sin(p_y a)\sin\varphi] \sin(p_z c),\eeq
where $g_T$ involves the derivative of the transverse hopping
$t_X$ with respect to strain.  In the particular case of sound
polarization parallel to the $x$-axis, the nearest neighbor bonds
that give the leading contribution to the sound attenuation, are
determined by: ${\mathbf{R}}=\pm{\mathbf{a}} \pm {\mathbf{c}}$
(see Fig. \ref{vectors}(b)). Summation over these bonds yields
\beq\label{Gamma_T_x} \Gamma_{ep}^{(T)}{(\varphi=0,{\bf{p}})}=g_T
\sin(p_x a) \sin(p_{z}c). \eeq In a similar manner, for phonon
polarization oriented along $y$-axis, the vertex is
\beq\label{Gamma_T_y}
\Gamma_{ep}^{(T)}{(\varphi=\pi/2,{\bf{p}})}=g_T \sin(p_y a)
\sin(p_{z} c). \eeq

It is natural to expect that the transverse coupling $g_T$ is
significantly less than the longitudinal one $g_L$. Still, it can
provide a sizable effect, according to preliminary experiments in
organics that suggest that the mechanism of the sound attenuation
has electronic origin at $T\rightarrow T_c$. At the same time, we
neglect the contributions from all further neighbor bonds that lie
for example along the spatial diagonals of the 3D unit cell,
assuming that the magnitudes of corresponding hopping integrals
decay fast enough.

Finally, note that the single-particle spectrum Eq.
(\ref{qparticle_spectrum_tight}) should be modified because of the
diagonal hopping $t_X$  mentioned above. That will introduce the
term given by Eq. (\ref{xi_X}). However, we will see in the next
section that this effect leads to small corrections, even in the
Aslamazov-Larkin diagram, where it is potentially important,
because $t_X/t_\perp\ll 1$.

\section{Ultrasound attenuation and sound velocity\label{sec_no_fluctuations}}

\subsection{Normal state\label{subsec_normal_state}}

We start with the sound attenuation and sound velocity in the
normal state. The sound attenuation coefficient is determined by
the imaginary part $ \gamma (k)$ of the complex frequency $\omega
(k)$ where the pole of the phonon Green function
$D(\mathbf{k},\omega )$ is located. This quantity obeys Dyson's
equation \cite{AGD}
\begin{equation}
D^{-1}(\mathbf{k},\omega _{\nu })=\left[ D^{0}(\mathbf{k},\omega _{\nu
})\right] ^{-1}-\Pi (\mathbf{k},\omega _{\nu }).  \label{Dyson_eq}
\end{equation}
Expressed in bosonic Matsubara frequencies $\omega _{\nu }=2\pi
\nu T$ using units $k_{B}=1$ $\hbar =1$, with $\nu $ an integer,
the quantity $D^{0}( \mathbf{k},\omega _{\nu })$
\begin{equation}
D^{0}(\mathbf{k},\omega _{\nu })=-\frac{\omega _{0}^{2}(\mathbf{k})}{\omega
_{\nu }^{2}+\omega _{0}^{2}(\mathbf{k})}  \label{Phonon_propagator}
\end{equation}
is the phonon propagator in the non-interacting case, and
 $\Pi (\mathbf{k},\omega _{\nu})$ is the phonon self-energy:
\begin{eqnarray}\label{Pi_NS}
&&\Pi (k,\omega _{\nu }) \\
&&{\!\!\!\!\!\!\!\!\!\!\!\!\!}=2T\sum_{\varepsilon _{n}}\int \frac{
d^{3}p}{(2\pi )^{3}}\Gamma_{ep} ^{2} G(\varepsilon
_{n},p)G(\varepsilon_{n}+\omega _{\nu },p+k).  \nonumber
\end{eqnarray}
In Eq. (\ref{Pi_NS}) the electron Green function
$G(\mathbf{p},\varepsilon _{n})$ at finite temperature in the
presence of impurities, is given by
\begin{equation}\label{electronic_GF}
G(\mathbf{p},\varepsilon _{n})=\frac{1}{i\tilde{\varepsilon}_{n}-\xi (
\mathbf{p})},
\end{equation}
where $\xi (\mathbf{p})$ is the quasiparticle energy and where we defined $
\tilde{\varepsilon}_{n}=\varepsilon _{n}+1/(2\tau )\mathrm{sign}(\varepsilon
_{n})$ with $\varepsilon _{n}=\pi T(2n+1)$ the Matsubara frequency and $\tau
$ the electronic elastic scattering time. The quantity $i/(2\tau )\mathrm{sign}%
(\varepsilon _{n})$ is the imaginary part of the quasiparticle
self-energy. The real part of the self-energy is constant and is
absorbed in the definition of the chemical potential.\ \cite{AGD}

The scattering time $\tau$ in Eq. (\ref{Pi_NS}) is the only effect
of impurity averaging in the polarization operator. We do not
include the corrections to the electron-phonon vertices
$\Gamma_{ep}$. In the case of the current-current correlator,
vertex corrections lead to the replacement of the scattering time
$\tau $ by its transport analog $\tau_{tr}$.\ \cite{AGD} For
$s$-wave scattering, vertex corrections vanish because the vector
vertex $ev_{\alpha }$ averages to zero upon angular integration at
vanishing external momentum $\mathbf{k}$. For stress tensor
correlator, that we need
in tight-binding limit, one can use results of Schmid \cite%
{Schmid}. He showed that in the continuum limit, taking into
account perfect screening, there is no impurity diffusion
enhancement of the electron-phonon vertex in the case of
transverse phonons, and that for longitudinal phonons this effect
is negligible in the hydrodynamic limit $k \ell \ll 1$ and $\omega
\tau \mathbf{\ll }1$. Physically, this comes about from the fact
that the calculation should be done in the moving frame and that
screening is perfect at long wavelengths. In our case, the
analogous argument of electroneutrality leads
\cite{SamokhinWalker} to the replacement of the stress vertex $F$ by $%
F-\left\langle F\right\rangle $, where the average accounts for
the chemical potential shift.\ \cite{Abrikosov} That average is
precisely what is needed to make the impurity vertex correction
vanish. In addition, in our specific case, $\left\langle
F\right\rangle   =0$ to the order in phonon wave vector $k$ that
we need.

As usual, after integration over $p$ and summation over $\eps_n$
in Eq. (\ref{Pi_NS}), one should make the analytic continuation of
the external phonon frequency following the rule $i\omega _{\nu
}\rightarrow \omega+i\delta $. For sound propagating along
$z$-axis, in the hydrodynamic limit $\omega \tau \ll 1$, it
suffices to set $k=0$ and expand the integrals in powers of
$\omega \tau$. The subtleties and the typical calculation details
are give in Ref. \cite{MBT04}, to which we will refer in what
follows.

The power attenuation can be obtained from
\begin{equation}
\alpha (\omega )=-2\gamma (\omega )/v_{s},  \label{alpha_general}
\end{equation}
where
\begin{equation}
\gamma (\omega )=\frac{1}{2}\omega _{0}(k)\;\mathrm{Im}\left[ \Pi^{R}(\omega
)\right] ,  \label{gamma}
\end{equation}
and $v_{s}$ is the sound velocity.

The renormalization of the phonon frequency $\omega (k)$ is obtained from
the real part of phonon self-energy using
\begin{equation}
\omega (k)= \omega_{0}(k)\sqrt{1+\mathrm{Re}\Pi ^{R}}.  \label{omaga_renorm}
\end{equation}

For both longitudinal and transverse sound polarizations, one finds:
\bea\label{sound_uni}
&&\alpha(\omega,\widehat{\mathbf{e}})=\frac{g_{\hat{e}}^2 \nu_0 \omega^2 \tau}{v_s},\\
&&v_s(\omega,\widehat{\mathbf{e}})=-g_{\hat{e}}^2 \nu_0,\nonumber
\eea where index $\hat{e}$ denotes the phonon polarization (T or
L). One can see that for transverse polarization, the phonon
self-energy does not depend on the orientation of the vector
$\widehat{\mathbf{e}}$ in the plane. This is true for both
tight-binding form of the spectrum Eq.
(\ref{qparticle_spectrum_tight}) and low-density limit Eq.
(\ref{qparticle_spectrum_lowdensity}). In the latter case, one can
analytically obtain: \beq\label{analytics_isotropy} \int_0^{2 \pi}
\left(\Gamma_{ep}^{(T)}\right)^2 d \varphi_p \sim \int_0^{2 \pi}
\cos^2 (\varphi-\varphi_p) d \varphi_p , \eeq that does not depend
on $\varphi$.

\subsection{Sound attenuation at $T<T_c$\label{subsec_belowTc}}

Before analyzing the fluctuation effects, let us consider the
effect of the in-plane phonon polarization and order parameter
symmetry on sound attenuation at temperature quite below $T_c$. It
is possible to show \cite{SamokhinWalker} that the sound
attenuation in the Born limit is given by:\beq\label{att_dwave}
\frac{\alpha(T)}{\alpha(T_c)} =\int_0^\infty  dE
\left(-\frac{\partial f }{\partial E}\right) \frac{A(E)}{E}, \eeq
where \beq\label{att_dwave_average} A(E)=\frac{\langle
\Gamma_{ep}^2(\varphi) \mbox{Re} \sqrt{E^2-|\Delta_p|^2}
\rangle_{F.S.}}{\langle \Gamma_{ep}^2(\varphi) \rangle_{F.S.}},
\eeq  with $\Delta_p$ the superconducting energy gap,
$E=\sqrt{\xi^2+|\Delta_p|^2}$, and $f(E)$ the Fermi distribution
function. For longitudinal phonons propagating in $s$- and
$d$-wave superconductor, $\Gamma_{ep}$ does not depend on
$\varphi$. For transverse phonons in $s$-wave superconductor the
attenuation does not depend on the polarization direction because
the energy gap $\Delta_p$ is isotropic in the plane, and because
of Eq. (\ref{analytics_isotropy}). Of course, the dependence of
the attenuation on temperature can be different for different
order parameter symmetry.\ \cite{SamokhinWalker, Contreras04,
ColemanMoreno}

\begin{figure}[h]
\includegraphics[width=5.0cm]{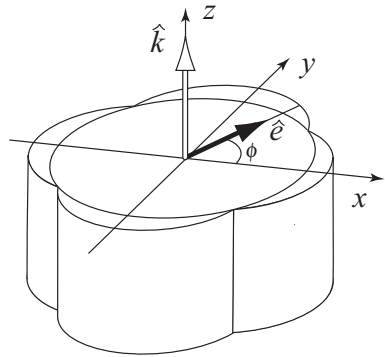}
\caption{Sketch showing the $d$-wave superconducting energy gap.
The ultrasound is propagating along $\hat{\bf{k}}$ direction with
the polarization $\widehat{\mathbf{e}}$ having an angle $\varphi$
with the $x$ axis.} \label{d-wave_gap_polarization}
\end{figure}

Let us consider the fourth case, namely, transverse phonons in
$d$-wave superconductor, at the energy gap given by
$\Delta_p=\Delta_0 \cos(2 \varphi_p)$. Let phonons propagate
perpendicular to the conduction layers (along the $\hat{z}$ axis,
Fig. \ref{d-wave_gap_polarization}), with the angle $\varphi$
between the phonon polarization vector $\widehat{\mathbf{e}}$ and
the $x$ axis in plane.
The electron-phonon vertex is then given by Eq.
(\ref{dt_elph_vertex}). As was pointed out in Sec.
\ref{subsec_normal_state}, in the normal state the sound
attenuation and sound velocity do not depend on the orientation of
the polarization in plane. Combining the expressions Eqs.
(\ref{dt_elph_vertex}), (\ref{att_dwave}), and
(\ref{att_dwave_average}), one can see that the sound attenuation
does not depend on the polarization direction below $T_c$ as well.
This can be demonstrated analytically in low-density limit using
Eq. (\ref{analytics_isotropy}), and numerically in more general
case of a square quasi-2D lattice. This should be contrasted with
the results of Refs. \cite{SamokhinWalker} and
\cite{ColemanMoreno}, where the attenuation of transverse phonons
with both $\widehat{\mathbf{k}}$ and $\widehat{\mathbf{e}}$ lying
in the plane does depend on the propagation direction.

\section{Fluctuation corrections, {\normalsize {S}}-wave superconductor\label{sec_fluct_s-wave}}

\subsection{Generalities\label{fl_generalities}}

\begin{figure}[h]
\includegraphics[width=8.0cm]{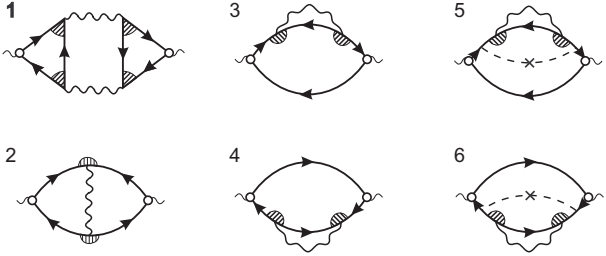} 
\caption{Feynman diagrams that give the leading-order corrections
from the superconducting $s$-wave fluctuations to the sound
attenuation as $T\rightarrow T_c$. Diagram 1 is of the
Aslamazov-Larkin (AL) type, diagram 2 is of Maki-Thompson (MT)
type, and the diagrams 3-6 are of the density of states (DOS)
type. Solid lines are the normal state Green functions, wavy lines
the fluctuation propagators, shaded semicircles  the impurity
ladder averaging, dashed lines with cross the single impurity
scattering, and open circles the renormalized electron-phonon
vertices. The corresponding diagrams with a $d$-wave order
parameter symmetry do not contain impurity semicircles (See Ref.
\cite{d-wave}).}\label{fl_diagrams}\end{figure}

The Feynman diagrams that give the main contribution to the
renormalization of the electron-phonon loop Eq. (\ref{Pi_NS}) by
superconducting fluctuations are presented in Fig.
\ref{fl_diagrams}.
 Here, each wavy
line corresponds to the fluctuation propagator (Cooper ladder)
$L(q,\Omega _{k})$ which, as $T\rightarrow T_{c}$, has the form
\cite{LV01}:
 \beq\label{L_s-wave}\!\!\!\!\!\!L^{(s)}(q,\Omega _{k})^{-1}=-\nu
_{0}\left[ \epsilon +\eta q_{||}^2 + r \sin^2(q_z c/2)+ \Omega_k
\tau_s \right],\eeq where $\epsilon \approx (T-T_{c})/T_{c}$,
$\Omega_k$ is the bosonic Matsubara frequency, and the coefficient
$\eta $ has the meaning of the square of the effective coherence
length $\xi $ in the isotropic 2D case for  $s$-wave pairing:\
\cite{LV01} \bea\label{eta_swave} &&\eta \equiv \xi ^{2}\left(
T\tau \right) =-\frac{\left( \tau v_{F}\right)
^{2}}{2} \\
&&\times \left. \left[ \psi \left( \frac{1}{2}+\frac{1}{4\pi x}\right) -\psi
\left( \frac{1}{2}\right) -\frac{1}{4\pi x}\psi ^{^{\prime }}\left( \frac{1}{%
2}\right) \right] \right\vert _{x=T\tau },  \nonumber \eea where
 $\psi (z)$ is the digamma
function. Then,  $r=16t_{\perp }^{2}\eta /v_{F}^{2}\ll 1$ is the
anisotropy parameter\ \cite{r}. At $T=T_{c}\,,$ the anisotropy
parameter can be written as $r=4\xi _{\perp }^{2}(0)/c^{2}$ where
$\xi _{\perp }(0)$ is the Cooper pair size in the perpendicular
$\left( z\right) $ direction. Finally, $\tau_s=\pi/(8 T)$ is the
Ginzburg-Landau time.

Returning to Fig.\ \ref{fl_diagrams}, the shaded semicircles correspond to
vertex corrections from impurity averaging and are given by:
\begin{equation}  \label{lambda}
\lambda (\mathbf{q},\varepsilon _{1},\varepsilon _{2})=\frac{|\tilde{
\varepsilon}_{1}-\tilde{\varepsilon}_{2}|}{|\varepsilon _{1}-\varepsilon
_{2}|+\frac{\widehat{D}q^{2}}{\tau ^{2}|\tilde{\varepsilon}_{1}-\tilde{
\varepsilon}_{2}|^{2}}\Theta (-\varepsilon _{1}\varepsilon _{2})}.
\end{equation}
We neglect the contribution from the diagrams that contain the
impurity ladder in the particle-particle channel \cite{MBT04}
because they are less singular in $\epsilon$ (they are not shown
on Fig. \ref{fl_diagrams}).

The open circles at the extreme left and right-hand sides of the
diagrams in Fig.\ \ref{fl_diagrams} represent the electron-phonon
vertices $\Gamma_{ep}({\bf{p}})$, that contain the dependence on
electron momentum and that will be different for longitudinal and
transverse phonon polarizations. Note that there is no impurity
averaging of the vertices in fluctuation diagrams just like in the
normal state polarization operator in the hydrodynamic limit.\
\cite{Schmid}.

As usual, the diagrams in Fig.\ \ref{fl_diagrams} correspond to
three different manifestations of superconducting fluctuations.\
\cite{LV01} a) Some of the electrons behave like Cooper pairs for
a time given by the Ginzburg-Landau time. This is the famous
Aslamazov-Larkin  (AL) contribution Fig.\ \ref{fl_diagrams}(1). b)
The single-particle excitations are Andreev reflected off the
superconducting fluctuations, as described by the so-called
Maki-Thompson (MT) term Fig.\ \ref{fl_diagrams}(2). c) The
effective number of normal carriers is reduced because some of the
electrons exist as transient Cooper pairs. This is the so-called
density of states (DOS) contribution in Fig.\ \ref{fl_diagrams}
(3-6). Additionally, the analytical expression for the  MT term
can be separated into the anomalous (aMT) and regular (rMT) parts.
The difference between them comes from the fact that there is
additional diffusion pole in the integral for the anomalous MT
part, that in most cases enhances the aMT fluctuation
corrections.\ \cite{LV01}

The evaluation of the integrals corresponding to the diagrams in
Fig. \ref{fl_diagrams}  includes taking the limit $\Omega_k=0$ in
the Green's functions and also in the fluctuation propagator of
the DOS and MT diagrams, and analytically continuing the diagrams
to real phonon frequencies $\omega_\nu \rightarrow - i \omega$. In
the hydrodynamic limit, where the electron mean free path $\ell $
is much smaller than the wavelength of sound and where the
electron collision rate $\tau ^{-1}$ is much larger than the sound
frequency, we expand the result in powers of $\omega \tau$ and we
set the phonon momentum $k =0$ because the expansion in powers of
$k \ell$ gives corrections negligible to leading order of $\omega
\tau $. The terms with odd powers of $\omega$ contribute to the
sound attenuation while the terms proportional to even powers of $
\omega $ contribute to the sound velocity.  In the most general
form the corrections can be written as:\bea\label{att_gen_form_s}
&&\!\!\!\!\!\!\!\!\!\!\!\!\!\!\!\!\!\!\!\!\!\Delta\alpha
^{(\beta,{\hat{e}},s )}(T,\omega )=\frac{g_{\hat{e}}^2 \omega
^{2}\nu _{0}}{\varepsilon_{F}v_{s}}
\kappa _{\alpha}^{(\beta,{\hat{e}},s)}(T\tau
)f^{(\beta,s)}_\alpha(\epsilon ,r,\gamma _{\phi }),\eea
\begin{eqnarray}\label{v_gen_form_s}
&&\frac{\Delta \omega ^{(\beta,{\hat{e}},s )}(T,\omega )}{\omega
}=\frac{\Delta v_{s}^{(\beta,{\hat{e}},s )}\left( T,\omega \right)
}{v_{s}}
\\
&&~~~~~~~~~~~~ =\frac{g_{\hat{e}}^2 T\nu _{0}}{\varepsilon
_{F}}\kappa _{v}^{(\beta,{\hat{e}},s)}(T\tau )f^{(\beta,s
)}_v(\epsilon ,r,\gamma _{\phi }). \nonumber
\end{eqnarray}
Here $\beta $ denotes the particular channel (DOS, rMT, aMT or
AL),  superscript ${\hat{e}}$ the phonon polarization (L or T).
The superscript $s$ stands for the $s$-wave symmetry of the order
parameter and in following Sections on $d$-wave symmetry it can be
also $d$. The symbol $f^{(\beta,s )}(\epsilon ,r,\gamma _{\phi })$
denotes the function of temperature which usually contains the
main singularity and comes from the integration over the momentum
${\bf{q}}$ in the fluctuation propagator Eq. (\ref{L_s-wave}), and
which also depends on the material properties. The $\gamma _{\phi
}=2\eta /(v_{F}^{2}\tau \tau _{\phi })$ in the temperature
functions is the cutoff parameter that appears in the anomalous MT
integrals and that depends on the phase-breaking time $\tau _{\phi
}$.\ \cite{Thompson} When the subscripts $\alpha$ or $v$ on
$f^{(\beta,s )}(\epsilon ,r,\gamma _{\phi })$ are omitted it is
because the function is identical for the attenuation and sound
velocity cases.

\subsection{Longitudinal phonons\label{subsec_Ls}}

For $s$-wave longitudinal case, the analytical expressions for the
temperature functions $f^{(\beta,s)}$ read:\ \cite{MBT04} \bea
&&\!\!\!\!\!\!\!\!\!\!\!\!f^{(DOS,s)}(\epsilon,r)=\ln \left(
\frac{2}{\sqrt{\epsilon }+\sqrt{
\epsilon+r }}\right),\label{temp_functions_s_DOS}\\
&&\!\!\!\!\!\!\!\!\!\!\!\!f^{(rMT,s)}(\epsilon,r)=
\frac{\left(\sqrt{\epsilon}-\sqrt{\epsilon+r}
\right)^2 }{r},\label{temp_functions_s_rMT}\\
&&\!\!\!\!\!\!\!\!\!\!\!\!f^{(aMT,s)}(\epsilon,r,\gamma_\phi)\\
&&\!\!\!\!\!\!\!\!\!\!\!\!~~~~~~~~=\frac{1}{r} \left[\frac{
\epsilon+r+\gamma_\phi}{\sqrt{\gamma_\phi(\gamma_\phi+r)}+ \sqrt{
\epsilon(\epsilon+r)}}-1 \right],\nonumber\label{temp_functions_s_aMT}\\
&&\!\!\!\!\!\!\!\!\!\!\!\!
f^{(AL,s)}_\alpha(\epsilon,r)=\frac{1}{r} \left[1-\sqrt{\frac{
\epsilon}{ \epsilon+r}} \left(1+\frac{r}{2(\epsilon+r)}\right)
\right],
\label{temp_functions_s_AL_att} \\
&&\!\!\!\!\!\!\!\!\!\!\!\!
 f^{(AL,s)}_v(\epsilon,r) = \left[1+ \frac{2 \epsilon}{r}
\left(-1+\sqrt{\frac{ \epsilon}{\epsilon+r}}\right)\right].
\label{temp_functions_s_AL_v}\eea The coefficients
$\kappa^{(\beta,{\hat{e}},s )}_\alpha(T\tau) $ in Eq.
(\ref{att_gen_form_s}) and $\kappa^{(\beta,{\hat{e}},s )}_v(T\tau)
$  in Eq. (\ref{v_gen_form_s}) come from the integration of Green
functions and impurity blocks. They have weaker temperature
dependence as $\epsilon \rightarrow 0$ than $f^{(\beta,s
)}(\epsilon ,r,\gamma _{\phi })$, and they basically show how the
impurity concentration affects the result. In Eqs.
(\ref{temp_functions_s_AL_att}) and (\ref{temp_functions_s_AL_v})
we introduced indices $\alpha$ and $v$ to distinguish between the
temperature corrections to the sound attenuation and the sound
velocity that are different in the AL channel.

The analytical form and an extended discussion of the asymptotics
of the $f^{(\beta,s )}(\epsilon ,r,\gamma _{\phi })$ functions and
the $\kappa$ coefficients can be found in Ref. \cite{MBT04}. In
short, the temperature functions experience sharp enhancement as
$\epsilon \rightarrow 0$, although they remain finite at $\epsilon
=0 $. Note that
$f^{(rMT,s)}(\epsilon=0,r)=f^{(AL,s)}_v(\epsilon=0,r)=1$, while
the value of functions $f^{(DOS,s)}(\epsilon,r)$,
$f^{(aMT,s)}(\epsilon,r,\gamma_\phi)$ and
$f^{(AL,s)}_\alpha(\epsilon,r)$ can be much larger than 1 at high
material anisotropy. In addition, long phase breaking times
increase $f^{(aMT,s)}(\epsilon,r,\gamma_\phi)$. Thus, with the
parameters corresponding to the real materials, the rMT diagram is
always negligible. In contrast to the conductivity fluctuations,
the contributions of the DOS diagram is enhanced and becomes
comparable to that of the AL and aMT diagrams, although its sign
is opposite to that of AL. The superconducting fluctuation
corrections to the sound attenuation, in a realistic range of
parameters, are given by the sum of the DOS, anomalous MT and AL
terms, which decrease the normal state attenuation. In contrast,
to leading order in $\omega $, the corrections to the sound
velocity are given by DOS diagrams, because the expansion for the
anomalous MT diagram begins at order $\omega $ while the AL
diagram is small to this order.

One can also see that the temperature dependence of the
microscopic terms Eqs. (\ref{temp_functions_s_DOS}),
(\ref{temp_functions_s_rMT}), and (\ref{temp_functions_s_AL_v}) is
in agreement with the thermodynamic results Eqs.
(\ref{LD_DOS})-(\ref{LD_AL_tight}) that depend on the modulation
of the hopping integral.

Summarizing, the signs of the principal terms for $s$-wave
longitudinal case is as follows: \bea \Delta \alpha ^{(DOS,L,s)}
<0, &~~~&\Delta \alpha ^{(aMT,L,s)} <0,\\
\Delta v ^{(DOS,L,s)} >0,&~~~&\Delta \alpha ^{(AL,L,s)}
>0,\nonumber\eea and all other terms can be neglected.

\subsection{Transverse phonons\label{subsec_Ts}}

In the $s$-wave transverse case, the structure of the DOS and MT
terms is as follows: \bea\label{MT2_1}
&& \Pi^{(\beta,T,s)}(\omega _{\nu})  \\
&&~~~=2g_T^{2} T\sum_{\Omega _{k}}\int \frac{d^{3}q}{(2\pi )^{3}}
L^{(s)}(q,\Omega _{k})K^{(\beta,T,s)}(q,\Omega _{k},\omega _{\nu
}), \nonumber \eea where $K^{(\beta,T,s)}(q,\Omega _{k},\omega
_{\nu })$ is the "bubble" that contains electronic Green functions
Eq. (\ref{electronic_GF}),  impurity vertices Eq. (\ref{lambda}),
and electron-phonon vertices Eq. (\ref{dt_elph_vertex}). There is
no significant modification of the integrals in comparison with
the longitudinal polarization case. Indeed, the angular dependence
of the electron-phonon vertex part  $\left[
\Gamma_{ep}^{(T)}{(\varphi,{\bf{p}})}\right]^2$ in the {\em DOS
diagram} is averaged out at the Fermi-surface.

In the {\em Maki-Thompson diagram}, the term
$\Gamma_{ep}^{(T)}{(\varphi,{\bf{p}})}\Gamma_{ep}^{(T)}{(\varphi,{\bf{q}}-{\bf{p}})}$
after averaging over the direction of ${\bf {p}}$ provides an
additional term $\cos(q_x a)$ in the integrals over ${\bf {q}}$:
\begin{equation}
\int \frac{d^{3}q}{(2\pi )^{3}}\frac{\cos (q_{x} a) \cos (q_{z} c)}{\epsilon
+\eta q_{||}^{2}+r\sin ^{2}(q_{z}c/2)},  \label{RMT_qint}
\end{equation}
for the regular part, and
\begin{equation}
\int \frac{d^{3}q}{(2\pi )^{3}}\frac{\cos (q_{x} a) \cos (q_{z}c)}{\left(
\hat{D}
q^{2}+1/\tau _{\phi }\right) \left[ \epsilon +\eta q_{||}^{2}+r\sin
^{2}(q_{z}c/2)\right] }.  \label{AMT_qint}
\end{equation}
for the anomalous part of the diagram. (We set $\varphi=0$ from
now on, assuming the polarization is along the $x$-axis. The
generalization of our results to arbitrary polarization is
straightforward and does not change our main conclusions). At $a
p_F \ll 1$, one can neglect the $q_{x}$-dependence in $\cos (q_{x}
a)$ in the above integrals and immediately obtain the longitudinal
MT terms Eqs. (\ref{temp_functions_s_rMT}) and
(\ref{temp_functions_s_aMT}), with the properly defined transverse
coupling constant $g_T$.

The {\em Aslamazov-Larkin diagram} is given by:
\begin{eqnarray}\label{AL}
&&\Pi ^{(AL)}(\omega _{\nu })=-2g_T^{2}\ T\sum_{\Omega _{k}}\int
\frac{ d^{3}q }{(2\pi )^{3}}B^{2}(q,\Omega _{k},\omega _{\nu })
\nonumber
\label{AL_1} \\
&&~~~~~~~~\times L^{(s)}(q,\Omega _{k})L^{(s)}(q,\Omega
_{k}+\omega _{\nu }),
\end{eqnarray}
where $B(q,\Omega _{k},\omega _{\nu }) $ is the triangular block
in the diagram (1) on Fig. \ref{fl_diagrams}. As usual, at $q
\rightarrow 0$, which gives the main contribution from fluctuation
propagators, the Green function $G(q-p)$ in the B-block should be
expanded in powers of $q$:
\bea\label{Gmp_expansion}&&G(q-p)=G(-p)+G^2(-p)
\Delta\xi(q,p)\\
&&~~~~~~~~~~~~+G^3(-p) \Delta\xi(q,p)^2+\ldots\nonumber\eea where
$\Delta\xi(q,p)=\xi(q-p)-\xi(p)$. One can see that with the
quasiparticle spectrum Eq. (\ref{qparticle_spectrum_lowdensity})
one obtains \bea\label{delta_xi_perp}
&&\Delta\xi(q,p)=\frac{q_x(q_x-2p_x)}{2m} \\
&&~~~~~~~~~~~~~~~+2 t_\perp \{ \cos [(q_z-p_{z})c]-\cos
(p_{z}c)\}.\nonumber\eea The first term of the expansion Eq.
(\ref{Gmp_expansion}) that does not vanish upon angular averaging
is  $G(-p)^3 (\Delta\xi(q,p))^2$. Thus the B-block contains a
product of {\it five} Green functions; one can check that this
does not lead to any singularity in the total AL diagram
contribution.

We must also consider the term in the quasiparticle energy
spectrum that depends on hopping $t_X\ll t_\perp$ along the
diagonal lattice bonds $\bf{R}=\pm \bf{a} \pm \bf{c}$ (See Fig.
\ref{vectors}). This term can be chosen as:
\beq\label{xi_X}\xi(p)_X=t_X \sin(p_x a) \sin(p_z c),\eeq such
that\bea\label{delta_xi_cross}&&\!\!\!\!\!\!\!\!\!
\Delta\xi(q,p)_X=\frac{q_x(q_x-2p_x)}{2m}
\\&&\!\!\!\!\!\!\!\!\! + t_X \{\sin[(q_x-p_x)a] \sin[(q_z-p_z) c]-\sin(p_x a)
\sin(p_z c)\}.\nonumber\eea With this modification, the first
non-vanishing term of the expansion Eq. (\ref{Gmp_expansion})
appears to first order in $\Delta\xi(q,p)_X$. The integration in
the B-block now is reduced to  $\int \langle \Gamma_{ep} \Delta\xi
\rangle_{F.S.} G(p) G(p+k) G^2(-p) d\xi_p$; one can see that the
integration over $d\xi_p$ gives the same result both for
longitudinal and transverse polarizations. Indeed, there is no
angular dependence in the integral $\int G(p) G(p+k)G^2(-p)
d\xi_p$. Then, the average $\langle\Gamma_{ep}^{(T)} \Delta\xi_X
\rangle_{F.S.}$ contains a factor: \beq\label{B_trans_angular}g_T
t_X \frac{\pi^2}{c} [1-\cos(a q_x)\cos(c q_z)],\eeq while its
analog in the longitudinal case $\langle\Gamma_{ep}^{(L)}
\Delta\xi_{\perp} \rangle_{F.S.}$ is:
\beq\label{B_long_angular}g_L t_\perp \frac{\pi^2}{c} [1-\cos(c
q_z)].\eeq Integrating over $q$ in Eq.  (\ref{AL}) with
 Eq. (\ref{B_trans_angular}) one can set $\cos(a q_x)\approx 1$, thus
reducing it to Eq.  (\ref{B_long_angular}). From Eq.  (26) of Ref.
\cite{MBT04}, one can easily obtain the expression for the
B-block:\begin{equation}\label{B_long}B(q)=-\frac{t_{X }\eta \nu
_{0}[1-\cos (c q_{z})]}{v_{F}^{2}}.\end{equation}We see that the
difference between the longitudinal and transverse sound
attenuation and velocity in AL diagram is only in the prefactor.
However, compared with the DOS and MT contributions, this
prefactor contains an additional $(t_X/t_\perp)^2$ term because of
the anisotropy parameter $r$ which enters the fluctuation
propagator which is still proportional to $t_\perp^2$. One would
expect that this prefactor satisfies $(t_X/t_\perp)^2\ll 1$, thus
making the AL term negligible.

In summary, the main contribution to the fluctuation corrections
is given by the DOS diagrams (and aMT diagrams in sound
attenuation at sufficiently long pair-breaking times). The
temperature function of the rMT diagram is less singular and the
AL diagram contains an additional small factor, thus being
negligible. The signs of the principal terms are as follows:
\bea\Delta \alpha ^{(DOS,T,s)} <0, &~~~&\Delta \alpha ^{(aMT,T,s)}
<0,\\ \Delta v ^{(DOS,T,s)}
>0.&~~~&\nonumber\eea

\section{Fluctuation corrections, {\normalsize $d$}-wave\label{sec_dwave}}

\subsection{Modification of the {\normalsize $d$}-wave integrals\label{subsec_dwave_modifications}}

If we assume that the virtual Cooper pairs have $d$-wave symmetry, the
modification of the calculations is as follows.

 We have to take into account the momentum dependence of a
pairing interaction that now  can be written as:
\beq\label{V_dwave} V(p,p^{\prime})=\eta(\varphi_p) g
\eta(\varphi_{p^{\prime}}), \eeq where, for $d$-wave pairing,
$\eta(\varphi_p)\sim\cos(a p_x)-\cos(a p_y) \propto
\sqrt{2}\cos(2\varphi_p)$.

Then, in all fluctuation diagrams of Fig. \ref{fl_diagrams}, each
wavy line corresponds to the fluctuation propagator with
additional symmetry factors: \beq\label{L_dwave_general}
 \eta(\varphi_p) L^{(d)}(q,\Omega_k)
\eta(\varphi_{p^{\prime}}), \eeq where $L^{(d)}(q,\Omega_k)$ has
the same form as the $s$-wave fluctuation propagator Eq.
(\ref{L_s-wave}) with modified $\epsilon$, $\xi$, $\tau$, and $r$:
\beq\label{Ld_q2D} L_d(q,\Omega)^{-1}=-\nu_0
\left[\epsilon_d+\eta_d q_{||}^2+r_d \sin^2 (c q_z/2)-i
\Omega\tau_d \right].\eeq In the quasi-2D case, and assuming that
the phase-breaking processes contribute only as the upper cutoff
for the Maki-Thompson diagram, the coefficients in  Eq.
(\ref{Ld_q2D}) read (compare with those of Ref. \cite{Sauls_tau}):
\bea\label{epsilon_d}&&\epsilon_d=\ln \frac{T}{T_c}
-\psi\left(\frac{1}{2}+\frac{1}{4 \pi  T_c \tau}
\right)+\psi\left(\frac{1}{2}+\frac{1}{4 \pi  T \tau}
\right)\nonumber\\&&~~~~~~~~  \approx \epsilon\left[ 1-\frac{1}{4
\pi x}\psi^\prime\left(\frac{1}{2}+\frac{1}{4 \pi  x} \right)
\right]
\\&&~~~~~~~~~~~~~~\approx \epsilon \left\{\begin{array}{lll}&  13.1595 x^2 , & ~~x\ll
1, \\&  &  \\& 1, & ~~x\gg 1;
\end{array}
\right.\nonumber
\eea
\bea\label{xi_d}
&&\eta_d= \left(\frac{v_F \tau}{8 \pi x}\right)^2 \left|~
\psi^{\prime\prime}\left(\frac{1}{2}+\frac{1}{4 \pi  x} \right)
\right|
\\&&~~~~~~~~~~~~~~
\approx \left(\frac{v_F }{2 T}
\right)^2\left\{\begin{array}{lll}& x^2, & ~~x\ll 1, \\
&  &  \\
& \frac{7\zeta(3)}{8 \pi^2 }, & ~~x\gg 1;
\end{array}
\right.\nonumber
\eea
\bea\label{tau_d}
&&\tau_d = \tau \frac{1}{4 \pi x}\psi^\prime\left(\frac{1}{2}+\frac{1}{4 \pi  x}
\right)
\\&&~~~~~~~~~~~~~~
\approx \frac{1}{T} \left\{
\begin{array}{lll}
&  x, & ~~x\ll 1, \\
&  &  \\
& \frac{\pi}{8}, & ~~x\gg 1;
\end{array}
\right.\nonumber
\eea
\bea\label{r_d}
&& r_d =
\left(
\frac{t_{\perp}}{4 \pi T}
\right)^2
\left|\psi^{\prime\prime}\left(\frac{1}{2}+\frac{1}{4 \pi  x} \right)
\right|
\\
&&~~~~~~~~~~~~~~\approx \frac{t_{\perp}^2 }{T^2}\left\{
\begin{array}{lll}&  x^2, & ~~x\ll
1, \\&  &  \\
& \frac{7 \zeta(3)}{8 \pi^2 }, & ~~x\gg 1.
\end{array}
\right.\nonumber \eea where $x=T \tau$. Note that the dependence
of $\eta_d$ and $r_d$ on the impurity concentration (through the
parameter $T\tau$) is the same. It also should be noted that the
$d$-wave parameters $\epsilon_d$, $\eta_d$, $r_d$, and $\tau_d$
coincide with the corresponding $s$-wave terms in the clean limit
($T\tau\gg 1$). In the opposite, dirty limit, the pair-breaking
effect of impurities for $d$-wave pairing makes these coefficient
different from those for $s$-wave pairing.

Note also that in the expression Eq. (\ref{epsilon_d}) $T_c$ is
the real superconducting transition temperature. In the case of
$d$-wave pairing, it is renormalized even by scattering on
non-magnetic impurities, and it satisfies the Abrikosov-Gorkov
equation:
\[
\ln\frac{T_c}{T_{c0}} -\psi\left( \frac{1}{2}\right) +\psi\left(
\frac{1}{2} +\frac{1}{4 \pi \tau T_c}\right)=0.
\]
In the clean case, at $\tau T \approx\tau T_c\gg 1$, there is not
much difference between $T_c$ and $T_{c0}$. In the opposite, dirty
limit, when the elastic scattering time is small, $d$-wave
superconductivity is suppressed and $T_c/T_{c0}\rightarrow 0$ as
$\tau T\rightarrow 0$. On the other hand, there is no such
impurity effect on the $s$-wave $T_c$.

Finally, there is no impurity averaging in the fluctuation
propagator (no shaded semicircles on diagrams Fig.
\ref{fl_diagrams}).\ \cite{d-wave} This also leads to the absence
of an anomalous part in the MT diagram 2 of Fig. \ref{fl_diagrams}
since there is no diffusion pole in the corresponding integral.

\subsection{Longitudinal phonons \label{subsec_Ld}}

\subsubsection{DOS}

The corrections to the sound attenuation and to the sound velocity
have the structure of Eqs.
 (\ref{att_gen_form_s}) and (\ref{v_gen_form_s}), with the temperature functions appropriate for $d$-wave pairing:
\begin{eqnarray}
\label{DOS_f_dl_alpha}&&f^{(DOS,d)}(\epsilon_d,r_d)=\ln \left(
\frac{2}{\sqrt{\epsilon_d }+\sqrt{ \epsilon_d+r_d }}\right) \\&&  \nonumber
\\&&~~~~~~~~~~\approx - \frac{1}{2}\left\{\begin{array}{lll}& \ln r_d\ ,\ ~~ &
\epsilon_d \ll r_d, \\&  &  \\
& \ln \epsilon_d\ ,\ ~~ & r_d \ll \epsilon_d,
\end{array}
\right.  \nonumber
\end{eqnarray}
with $\epsilon_d$ and $r_d$ given by Eqs. (\ref{epsilon_d}) and (\ref{r_d}). The
renormalization of $\tau_d$ Eq. (\ref{tau_d}) does not affect the results because the
main contribution from the fluctuation propagator is at $\Omega \rightarrow
0$.

The $\kappa^{(DOS,L,d )}$ coefficients  are inversely proportional
to the renormalized $\eta_d=\xi_d^2$  in Eq. (\ref{xi_d}) which is
different from its $s$-wave analog. Another source of difference
between $s$-wave and $d$-wave results is the absence of impurity
vertices $\lambda$ in $d$-wave diagrams. With all these
modifications, one finally obtains  the dependence of the
$\kappa^{(DOS,L,d )}$ coefficients on sample cleanness through the
parameter $T \tau \approx T_c\tau$.

For {\it sound
attenuation}, the coefficient is given by:
\begin{eqnarray}\label{DOS_kappa_dl_alpha}&&\kappa _{\alpha}^{(DOS,L,d)}(x)=
\frac{1}{\pi\psi^{\prime\prime}\left(\frac{1}{2}+\frac{1}{4\pi x}\right)}\\
&&\times\left[
32  \pi^2 x^2 \psi^{\prime}\left(\frac{1}{2}+\frac{1}{4\pi x}\right)
+ 8 \pi (2\pi-1)x
\psi^{\prime\prime}
\left(\frac{1}{2}+\frac{1}{4\pi
x}\right)\right.\nonumber\\
&&~~~~~~~~~~~~~~~~~~~~\left.
+\psi^{\prime\prime\prime}\left(\frac{1}{2}+\frac{1}{4\pi
x}\right)
\right]
\nonumber\\
&&~~~~~~~~~~\approx \left\{
\begin{array}{lll}
& 4 (2 \pi-3)x, & ~~x\ll 1,\\
&  &  \\
& -{\mbox{\Large $\frac{4 \pi^3 x^2}{7 \zeta(3)}$}}, & ~~x\gg 1,
\end{array}\right.\end{eqnarray}
while for {\it sound velocity} it reads:
\beq\label{DOS_kappa_dl_v} \kappa_{v}^{(DOS,L,d)}=1. \eeq

The temperature function $f^{(DOS,d)}(\epsilon_d,r_d)$ Eq.
(\ref{DOS_f_dl_alpha}) is similar to its $s$-wave analog Eq.
(\ref{temp_functions_s_DOS}), although now it also depends on the
impurity concentration. Qualitatively, it is monotonically
increasing as $T\rightarrow T_c$ just like
$f^{(DOS,s)}(\epsilon,r)$.

\subsubsection{rMT}Modifications similar to those
in the DOS integrals also occur in the MT diagram.   There is no aMT contribution because
there is no diffusion pole originating from the impurity vertices.
For rMT term, the temperature function is:
\begin{eqnarray}\label{rMT_f_dl_alpha}
&&f^{(rMT,d)}(\epsilon_d,r_d)=
\frac{\left(\sqrt{\epsilon_d}-\sqrt{\epsilon_d+r_d}\right)^2 }{r_d} \\
&&  \nonumber \\
&&~~~~~~\approx \left\{
\begin{array}{lll}
& 1\ ,\ ~~ & \epsilon_d \ll r_d, \\
&  &  \\
& {\mbox{\Large $\frac{r_d}{4 \epsilon_d }$}}\ ,\ ~~ & r_d \ll \epsilon_d.
\end{array}
\right.  \nonumber
\end{eqnarray}
Similarly to the $s$-wave case, the function in Eq.
(\ref{rMT_f_dl_alpha}) has a limit
$f^{(rMT,d)}(\epsilon_d=0,r_d)=1$. This function is very little
affected by $T\tau$. The $\kappa ^{(rMT,L,d)}$ coefficients are
finite (they do not contain a singularity as functions of $x$). In
summary, the rMT contributions to the attenuation and sound
velocity is negligible in comparison with those from the DOS
diagrams.

\subsubsection{AL \label{subsubsec_Ld_AL}}
The triangle of Green functions in diagram 1 in Fig.
\ref{fl_diagrams} does not contain impurity semicircles and reads:
\begin{eqnarray}\label{AL_Bblock_LD}
&&\!\!\!\!\!\!B^{(L)}_d(q,\Omega _{k},\omega _{\nu
})=T\sum_{\varepsilon _{n}}\int \frac{d^{3}p}{(2\pi )^{3}}  \eta^2
(\varphi_p) \Gamma^{(L)}_{ep}(p_z)
 \\&&\!\!\!\!\!\!   \times G(p,\varepsilon
_{n})G(p,\varepsilon _{n}+\omega _{\nu})G(q-p,-\varepsilon _{n}).
\nonumber
\end{eqnarray} One can see that
just like in the "longitudinal $s$-wave" case, it suffices to
expand the Green function $G(q-p,-\varepsilon _{n})$ to first
order in hopping $t_\perp$. The evaluation of Eq.
(\ref{AL_Bblock_LD}) with
$\eta(\varphi_p)=\sqrt{2}\cos(2\varphi_p)$ gives the same result
as in the case of $s$-wave pairing with $\eta(\varphi_p)=1$. Then,
the temperature functions of the AL diagram can be obtained from
those for the $s$-wave pairing Eqs.
(\ref{temp_functions_s_AL_att}) and (\ref{temp_functions_s_AL_v})
with the substitution of $\epsilon_s$, $\eta_s$, $\tau_s$, and
$r_s$ by their $d$-wave analogs Eqs.
(\ref{epsilon_d})-(\ref{r_d}).

In the clean limit we recover for $\kappa^{(AL,L,d)}_\alpha$ and
$\kappa^{(AL,L,d)}_v$ their longitudinal $s$-wave,
$T\tau$-independent analogs \cite{MBT04}. On the other hand,
$\kappa^{(AL,L,d)}$ coefficients remain finite in the dirty limit.
Moreover, in highly anisotropic materials at $r_d\ll \epsilon_d$,
the AL contribution to the sound attenuation is clearly
suppressed, and becomes important only at temperatures extremely
close to $T_c$, just like the corresponding $s$-wave terms. The AL
correction to the sound velocity is negligible just like the rMT
one.

\subsubsection{Summary}
In summary, in the $d$-wave fluctuation corrections to
longitudinal sound   the aMT term is absent, and rMT and AL sound
velocity corrections can be neglected in the range of parameters
that is reasonable. The signs of the principal terms are as
follows: \bea\Delta \alpha ^{(DOS,T,s)} <0, &~~~&\Delta \alpha
^{(AL,T,s)}
>0,\\ \Delta v
^{(DOS,T,s)}
>0.&~~~&\nonumber\eea
The magnitude of the corrections is finite though large in the
clean limit, and it is suppressed to zero in the very dirty limit.
Again, like in the $s$-wave case, the temperature dependence of
the microscopic expressions for $d$-wave order parameter symmetry
can be obtained in the phenomenological approach with the
substitution of proper $\eps_d$, $\eta_d$ and $r_d$ in the
fluctuation term Eq. (\ref{Omega_fl_q2D}) and taking into account
the strain dependence of $t_\perp$.

\subsection{Transverse phonons\label{subsec_Td}}

One can easily combine the results of Sec. \ref{subsec_Ls} and
Sec. \ref{subsec_Ld}.

The {\it DOS contribution} is given by Eq.  (\ref{MT2_1}) with the
fluctuation propagator modified according to
 Eq. (\ref{L_dwave_general}). It is clear that after such a substitution
one should obtain the results for the DOS longitudinal $d$-wave
term because the angular dependence in
$\left[\Gamma^{(T)}_{ep}\right]^2$ is averaged out.

Similar arguments are valid for the {\it rMT contribution}.  Indeed, the longitudinal
term contains the combination\bea
&&\Gamma^{(L)}_{ep}({\mathbf{p}}) \Gamma^{(L)}_{ep}({\mathbf{q-p}}) \\
&&=g_L^2 \cos(p_z c) \cos[(q_z-p_z)c].\nonumber \eea The
corresponding combination in the transverse term at, for example,
$\varphi=0$ can be rewritten to first non-vanishing order in the
low-density limit $a p_F \ll 1$:\bea
&&\Gamma^{(T)}_{ep}(\varphi=0,{\bf{p}}) \Gamma^{(T)}_{ep}(\varphi=0,{\bf{q-p}}) \\
&=& g_T^2 \sin (p_x a)  \sin (p_z c)  \sin [(q_x-p_x) a] \sin [(q_z-p_z)c]\nonumber \\
&\approx& \tilde{g}_T^2 \cos(q_x a) \cos^2(\varphi_p) \sin(p_z c)
\sin[(p_z-q_z)c],\nonumber \eea where $(a p_F)^2$ is absorbed in
$\tilde{g}_T$, and the term linear in $\cos(\varphi_p)$ drops out
after angular integration.

Since \bea &&\int_{-\pi/c}^{\pi/c} \cos(p_z c) \cos[(q_z-p_z)c]
dp_z\\&&~~~ = - \int_{-\pi/c}^{\pi/c} \sin(p_z c) \sin[(q_z-p_z)c]
dp_z=\frac{\pi }{c} \cos(q_z c)\nonumber, \eea one can see that
after integrating over $p_z$, the transverse term should contain
an extra $\cos(q_x a) \cos^2(\varphi_p)$ in comparison with the
longitudinal one. Then,  it suffices to set $\cos(q_x a)\approx 1$
at  $q_xa\ll 1$, and $\cos^2(\varphi_p)$ will provide an extra
$1/2$ after the angular integration. We conclude that with the
proper choice of the coupling constant, one can use the
longitudinal $d$-wave results.

For the same reason as in the longitudinal
$d$-wave case, there is no {\it aMT contribution} here.

Finally, as described in Sec. \ref{subsec_Ls}, in order to obtain
non-zero {\it AL contribution}, one should expand the Green
function $G(q-p)$ either to second order in small
$\Delta\xi(q,p)_{\perp}$  as in Eq. (\ref{delta_xi_perp}) (this
results in non-singular AL contribution), or to linear order in
$\Delta\xi(q,p)_{X}$ (this adds an additional small factor
$(t_X/t_\perp)^2$ in front of the whole diagram). The details of
the integration over $q$ are the same as in the transverse
$s$-wave case. We conclude that the AL term can be neglected too.

The leading terms then read: \bea\Delta \alpha ^{(DOS,T,s)} <0,
&~~~&\Delta v ^{(DOS,T,s)}
>0.\eea

\section{Discussion}

The superconducting fluctuations can provide a renormalization of
the normal state sound attenuation and sound velocity that depends
on the various phonon polarizations and order parameter
symmetries. The magnitude of the effect, as can be seen from Eqs.
(\ref{sound_uni}) and (\ref{att_gen_form_s}), is of the order of
$T_c/\varepsilon_F$ or less, depending on the material cleanness.
For layered organics, the Fermi surface parameters of Ref.\
\cite{Dressel} lead to $ T_c/\varepsilon _{F}\sim 10^{-2}$. The
temperature functions $f^{(\beta,s )}(\epsilon ,r,\gamma _{\phi
})$ increase this ratio at $T\rightarrow T_{c}$, thus making the
fluctuation corrections experimentally measurable.

The actual sign of the fluctuation corrections is determined by
the microscopic coefficients $\kappa
_{\alpha}^{(\beta,{\hat{e}},s)}$. We gather the information on
signs and relative magnitudes of the microscopic fluctuation terms
in Table I.
\begin{table*}[htbp]\label{tab_gather}
\begin{tabular}{|c | c|c|c|c|c|c|c | }
\hline\hline symmetry&mode&channel&&DOS&rMT&aMT&AL\\ \hline
\multirow{4}{*}{$s$} & \multirow{2}{*}{L[001]} &$\alpha$&&$-$&$\la
|+1|$&$-$&+\\
&&$v_s$&&+&$\la |-1|$&$O(\omega^2)$&$\la |-1|$ \\
&\multirow{2}{*}{T[001]} &$\alpha$&&$-$&$\la |+1|$&$-$&$O\left((t_X/t_\perp)^2\right)$\\
&&$v_s$&&+&$\la |-1|$&$O(\omega^2)$&$O\left((t_X/t_\perp)^2\right)$\\
\hline \multirow{4}{*}{$d$} & \multirow{2}{*}{L[001]}
&$\alpha$&&$-$&$\la |+1|$&none&+\\
&&$v_s$&&+&$\la |-1|$&none&$\la |-1|$ \\
&\multirow{2}{*}{T[001]} &$\alpha$&&$-$&$\la |+1|$&none&$O\left((t_X/t_\perp)^2\right)$\\
&&$v_s$&&+&$\la |-1|$&none&$O\left((t_X/t_\perp)^2\right)$\\
\hline\hline
\end{tabular}
  \caption{Summary of the fluctuation contributions from different diagrams for various
  order parameter symmetries and phonon polarizations. The symbol $\la |-1|$ means that
  the sign of the correction is negative and that the absolute value of the temperature function is less than unity
  with limit unity at $T=T_c$.}
 \end{table*}
 In short, the leading contribution is always in the DOS channel, and there are comparable corrections from the aMT
 diagram only for $s$-wave attenuation but for both polarizations, and from the AL diagram
 only for longitudinal sound attenuation but for both symmetries. The rMT
 (always) and the AL (longitudinal sound velocity) temperature
 functions   have a limit unity at $T=T_c$  but
 could compete with the leading terms only in the special cases of low
 anisotropy and short phase breaking time where the leading terms are suppressed. The aMT terms in the $s$-wave
 sound velocity and the AL terms for transverse phonon polarization
 contain additional small factors and can thus be neglected.
 Finally, there is no aMT term for $d$-wave symmetry of the order
 parameter because the diffusion pole disappears in this case.

The temperature dependence in the sound velocity obtained from the
phenomenological Ginzburg-Landau free energy is in agreement with
the microscopic calculations for the  longitudinal phonons in both
the $s$-wave and $d$-wave cases. In the latter case, it suffices
to replace $\epsilon$, $\eta$ and $r$ by their $d$-wave
equivalents $\epsilon_d$, $\eta_d$ and $r_d$ defined in Sec.
\ref{sec_dwave}. For example, one can compare the temperature
dependence of the phenomenological result Eq. (\ref{LD_DOS}) with
that of the DOS contribution in Eqs. (\ref{temp_functions_s_DOS})
and (\ref{DOS_f_dl_alpha}). Also, the temperature dependence of
the phenomenological Eq. (\ref{LD_rMT}) as well as the sign of the
correction is the same as that of the rMT terms Eqs.
(\ref{temp_functions_s_rMT}) and (\ref{rMT_f_dl_alpha}).
Similarly, the AL contributions (sign and temperature function)
Eqs. (\ref{temp_functions_s_AL_v}) and Sec. \ref{subsubsec_Ld_AL}
are explained by the phenomenological result Eq.
(\ref{LD_AL_tight}). The Ginzburg-Landau approach does not treat
the aMT contribution correctly because it does not include the
dynamics involved in the diffusion.

In principle, one can obtain the fluctuation corrections to the
sound velocity of transverse phonons from the Ginzburg-Landau
approach as well. The modification will include taking into
account the next-to-nearest neighbor hopping in the quasiparticle
energy spectrum Eq. (\ref{qparticle_spectrum_lowdensity}), extra
hopping terms in the fluctuation free energy Eq. (\ref{Omega_fl}),
and choosing the appropriate combination $\overline{c}$ in Eq.
(\ref{sound_velocity_density}) that will modify the elastic terms
Eq. (\ref{GammaDelta}).

 It should be noted that the relation $  {v_s}_{\mbox{fl}}
\propto -  C_{\mbox{fl}}$ between fluctuation corrections to the
sound velocity and specific heat that is valid in the bulk
continuous model, is not satisfied in the tight-binding model with
hopping terms in the electron quasiparticle energy spectrum. Thus,
the experimental information from the specific heat measurements
(first reference in \cite{fluct_ex}) is not enough to predict the
behavior of the sound velocity fluctuation corrections in organic
compounds.

We have obtained these results in the  quasi-2D square lattice
model for a corrugated cylindrical Fermi-surface. As a result, we
see, for example, that there is no effect on the attenuation of
the transverse sound as the direction of polarization is changed
in the plane, both in the normal and in the superconducting
states. In order to make a better fit of the experimental data,
the generalization of our model to the particular crystal lattice
(that is not tetragonal) and to the appropriate band structure is
important. For example, it might be necessary to incorporate the
dimer structure and triangular in-plane symmetry of organic
materials.\ \cite{Williams, Kino, McKenzie} That would better
reflect the two-band energy spectrum of the $\kappa$-(ET)$_2$X
compounds. All these problems should be the subject of future
research.

In conclusion, we have found how the sound attenuation and sound
velocity are renormalized by superconducting fluctuations at
temperatures close to $T_c$ in layered superconductors.  For
various polarizations of phonons propagating perpendicular to the
conduction plane, we considered the cases of $s$- and $d$-wave
symmetry of the order parameter. In the hydrodynamic limit $\omega
\tau \ll 1$ and $k\ell \ll 1 $, we found the contributions from
all the fluctuation diagrams (namely, the AL, MT and DOS) and
provided the theoretical background for the analysis of the
experimental information  in organic superconductors. The complete
analytical expressions for the microscopic fluctuation corrections
can be found elsewhere \cite{web_dwave}.

\section{Acknowledgements}

The authors would like to thank K.V.~Samokhin and M.~Walker for
valuable discussions, and D.~Fournier, C.~Lupien, M.~Poirier, and
L.~Taillefer for information on recent experiments in the field.
The present work was supported by the Natural Sciences and
Engineering Research Council (NSERC) of Canada, the Fonds de la
Recherche sur la Nature et les Technologies of the Qu\'{e}bec
government, and the Tier I Canada Research Chair Program
(A.-M.S.T.).

\end{document}